------------------------------------------------------------------------------
\date {}
\documentstyle{article}
\begin{document}
\large
\title  {\bf  A Study of Weak Mesonic Decays of $\Lambda_c$ and $\Xi_c$ 
Baryons on the  Basis of HQET Results}
\author{ K.K. Sharma and R. C. Verma$^*$ \\
\normalsize Centre for Advanced Study in Physics, Department of Physics, \\
\normalsize Panjab University, Chandigarh -160014, {\bf India}. }
\maketitle
\begin{abstract}
We investigate two-body Cabibbo-angle enhanced weak decays of charmed baryons
into octet baryon and pseudoscalar meson in the current algebra framework with
inclusion of the factorization terms which are evaluated using the HQET guided 
baryonic form factors.  We  obtain  the branching ratios  and  
asymmetry  parameters  for various Cabibbo-enhanced decays of $\Lambda_c^+$, 
$\Xi_c^+$ and $\Xi_c^0$ baryons. Sensitivity of the flavor dependence of the 
spatial wavefunction  overlap on the branching ratios and asymmetry parameters 
is also investigated. 
\end{abstract}
\vskip 0.5 cm
PACS number(s): 13.30.Eg, , 11.30.Hv, 11.40.Ha, 14.20.Kp \\
\vskip 1.0 cm
$^{*}$ Present Address: Department of Physics, Punjabi University, Patiala  - 147 002, India.
\newpage
\large
\section{Introduction} A significant progress in the experimental
determination of masses, lifetimes of charmed baryons and their decays
has taken place during the last few years. Masses of the charm unity
baryons have been measured within accuracy of a few percent. Charmed
baryons can decay through numerous channels. However, data on their
exclusive weak decays are available mainly for $\Lambda_c^+$ baryon [1],
though, a few decay modes of $\Xi_c^+$ baryon have also been observed [2].
Recently, the asymmetry parameters of $\Lambda_c^+ \rightarrow \Lambda 
\pi^+ $ and $ \Lambda_c^+ \rightarrow \Sigma^+\pi^0$ decays have been 
measured by the CLEO collaboration [3].
In the near future a large quantity of new and more accurate data on the
exclusive nonleptonic decays of heavy baryon can be expected which calls
for a comprehensive analysis of these decays. 

\par Even the meager data available for the charm baryon decays 
have already started to distinguish between various theoretical models. 
These models have been developed employing the flavor
symmetries [4], factorization [5], pole model [6-8], current algebra
[9,10] frameworks. So far none of these attempts has been able to explain 
the available data on the nonleptonic decays of the charmed baryons. 
The analysis of weak hadronic decays of baryons gets 
complicated by their being the three quark systems. Further, it not 
straightforward to estimate the strong interaction effects on their decays.
Initially, it was hoped that like
meson decays the spectator quark processes would dominate  charm baryon
decays also. However, this scheme does not seem to be supported by the
experiment as the observed branching ratios of
$\Lambda_c^+ \rightarrow \Sigma^+ \pi^0$, $\Lambda_c^+ \rightarrow 
\Sigma^0 \pi^+$, $\Lambda_c^+ \rightarrow \Sigma^+ \eta$ and 
$\Lambda_c^+ \rightarrow \Xi^0 K^+$ decays, forbidden by the spectator 
quark process, are significantly large
thereby indicating the need of the W-exchange contributions. Unlike the
mesons,  W-exchange seems to play a dramatic role in the charmed baryon decays,  
as this mechanism is neither  helicity nor color suppressed in baryon decays
due to the presence of of a scalar diquark system inside the baryons. 
Theoretically the contribution from this process has been expected to be 
proportional to $|\psi(0)|^2$, which renders it quite significant for 
these decays.

\par For two-body baryon decays, W-emission process leads to the
factorization which expresses decay amplitude as coupling of weak baryon
transition with the meson current. The matrix elements of the weak
transition between baryon states in general involve six form factors
which control the factorization contributions [5]. Fortunately, in the
past few years the discovery of new flavor and spin symmetries has
simplified the heavy flavor physics [11]. In the framework of Heavy
Quark Effective Theory (HQET), the form factors get mutually related,
though $1/M$ corrections are certainly needed [12,13]. At present one
does not know how to carry out these corrections from first principles
particularly for heavy to light baryon transitions and one takes the
help of phenomenological models. Recently, Cheng and Tseng [14] have
determined such corrections to the baryonic form factors in the
nonrelativistic quark model, which gives excellent agreement with the
experimental value for the only measured semileptonic decay $\Lambda_c^+
\rightarrow \Lambda e^+\nu_e$. Similar result has also been obtained by
Ivanov et al. in a relativistic three quark model [15]. It is worth
remarkable here that the agreement has been achieved due to the
flavor-suppression factor, resulting from the HQET considerations, for
the factorizable contribution. The full implications of this feature for
the nonleptonic decays of the charmed baryons is yet to be considered.
 
\par In the present work, we study Cabibbo-enhanced two-body weak
 decays of $\Lambda_c^+$, $\Xi_c^+$ and $\Xi_c^0$ 
into octet baryons $(J^P=1/2^+)$ and a
 pseudoscalar meson $(J^P=0^-)$. We include the factorization terms
 using the HQET guided form factors and the nonspectator
 contributions. Since current algebra is the most common approach used
 before for the study of the nonleptonic decays, we employ it to obtain
the nonspectator contributions. Section-II  describes the methodology of
 the calculations. Section-III deals with the numerical results for
 branching ratios and asymmetries of the charmed baryon decays and their
 comparison with the available data.  Here, we also study the effect of
 flavor dependence of the $|\psi(0)|^2$ on these  decays. In our
 analysis, we find that all factorization, pole and equal time current
 commutator (ETC) terms are equally important in the charm baryon
 decays, though, one may dominate over other depending upon the decay
 channel.
\section{Methodology} \subsection{Weak Hamiltonian} 
The general weak current $\otimes$ current weak Hamiltonian for Cabibbo
enhanced ($\Delta C= \Delta S = -1$) decays in terms of the quark fields is 
\begin{equation} H_{W}= {G_{F}\over {\sqrt 2}} V_{ud}V_{cs}^{*}[ c_{1}
(\bar u d) (\bar s c) + c_{2} (\bar s d) (\bar u c)], \label {(1a)}
\end{equation} where $\bar q_{1}q_{2} \equiv \bar
q_{1}\gamma_{\mu}(1-\gamma_{5})q_{2}$ represents the color-singlet
combination. $V_{ud}$ and $V_{cs}$ are the Cabibbo-Kobayashi-Maskawa (CKM)
weak mixing matrix elements. 
The perturbative QCD coefficients for the charm sector, $c_{1} 
= {1 \over 2}(c_{+} + c_{-}) = 1.26 \pm 0.04$ and $c_{2} = {1 \over 2}(c_{+}
- c_{-}) = -0.51 \pm 0.05$, are usually taken at the charm mass scale [16].
\subsection{Decay Width and Asymmetry Formulas} 
The matrix element for the baryon ${1 \over 2}^{+} \rightarrow
{1\over 2}^{+} + 0^{-}$decay process is written as 
\begin{equation} M =
- \langle B_{f} P \vert H_{W}\vert B_{i} \rangle = i \bar u_{B_{f}}( A -
\gamma_{5}B) u_{B_{i}} \phi_{P} ,\label {(5)} 
\end{equation} where A and
B represent the parity violating (PV) and the parity conserving (PC) amplitudes
respectively. The decay width is computed from 
\begin{equation} \Gamma =
C_{1} [ \vert A \vert^{2} + C_{2} \vert B \vert ^{2} ], \label {(6)}
\end{equation} where 
\begin{equation}C_{1} = \frac { \vert {\bf q}
\vert} {8 \pi} \frac { (m_{i} + m_{f})^{2} - m_{P}^{2}} {m_{i}^{2}},
\end{equation} 
\begin{equation}C_{2} = \frac { (m_{i} - m_{f})^{2} -
m_{P}^{2}}{(m_{i} + m_{f})^{2} + m_{P}^{2}}, 
\end{equation}
and 
\begin{equation}\vert {\bf q} \vert = \frac {1}{2m_{i}} {\sqrt
{[m_{i}^{2}- (m_{f} - m_{P})^{2}][m_{i}^{2} - (m_{f} +
m_{P})^{2}]}},
\end{equation} 
is the magnitude of centre  of mass
three-momentum in the rest frame of the parent particle.  $m_{i}$ and
$m_{f}$ are the masses of the initial and final baryons and $m_{P}$ is
the mass of the emitted meson. Asymmetry parameter is given by
\begin{equation} \alpha = \frac { 2 Re ( A \bar B^{*})}{ ( |A|^{2} +
|\bar B |^{2})}, \label {(7)}
\end{equation} where $\bar B = \sqrt
{C_{2}} B$.
\subsection { Decay Amplitudes}
Generalising  the current algebra (CA) framework of the hyperon decays [9,10]  to
the charm sector, the charmed baryon decay amplitudes 
receive contributions from the pole
diagrams involving the W-exchange process and the
ETC term. Including the factorization contributions, the nonleptonic
decay amplitude becomes \begin{equation}< B_f P | H_W | B_i > =
M_{fac} + M_{ETC} + M_{pole}. \label {(fep)} 
\end{equation}
We discuss the contribution of each of these terms in the context of PC
and PV amplitudes.
\subsubsection{Factorization Contributions}
The first term $M_{fac}$ in eq(8), corresponding to the factorization
contribution, can be obtained by inserting vacuum intermediate states,
which express it as a product of two current matrix elements [5];
\begin{equation}<P | A_{\mu} |0> <B_{f}(P_f )| V^{\mu} - A^{\mu} |
B_{i}(P_i) > \label {(f1)} 
\end{equation}  where 
\begin{equation}<P |
A_{\mu} |0> = i f_P q_{\mu} \label {(qm)} 
\end{equation} with $q_{\mu}$
being the meson four momenta, and $f_P$ is the decay constant of the
meson emitted. The matrix element for the baryonic transition $B_i
\rightarrow B_f$ can be expanded as 
\begin{equation}< B_{f}(P_f )|
V_{\mu} | B_{i}(P_i) > = \bar {u}_{f} (P_f) [ f_{1}(q^{2}) \gamma_{\mu}
+ i f_{2}(q^{2}) \sigma_{\mu \nu} q^{\nu} + f_{3}(q^{2}) q_{\mu} ] u_{i}
(P_{i}), \label {(f2)} 
\end{equation} 
\begin{equation}< B_{f}(P_f )|
A_{\mu} | B_{i}(P_i) > = \bar {u}_{f}(P_f) [ g_{1}(q^{2}) \gamma_{\mu} +
i g_{2}(q^{2}) \sigma_{\mu \nu} q^{\nu} + g_{3}(q^{2}) q_{\mu} ]
\gamma_{5} u_{i} (P_{i}), \label {(f3)} 
\end{equation}
where $f_i$'s and $g_i$'s are the vector and axial vector form factors. 
In the HQET framework, the matrix elements can be parameterised [11] 
in terms of the baryon velocities $v$ and $v'$,
\begin{equation}< B_{f}(v' )| V_{\mu} | B_{i}(v) > = \bar {u}_f [
F_{1}(\omega) \gamma_{\mu} + F_{2}(\omega) v_{\mu} + F_{3}(\omega)
v'_{\mu} ] u_{i} , \label {(f4)} 
\end{equation} 
\begin{equation}<
B_{f}(v' )| A_{\mu} | B_{i}(v) > = \bar {u}_f [ G_{1}(\omega)
\gamma_{\mu} + G_{2}(\omega) v_{\mu} + G_{3}(\omega) v'_{\mu} ]
\gamma_{5}u_{i}, \label {(f5)}  
\end{equation} with $ \omega = v \cdot
v'$. The form factors $f_i$'s and $g_i$'s are related to $F_i$'s and
$G_i$'s via 
\begin{equation}f_1 = F_1 + {1 \over 2 } (m_i + m_f) ( {
F_{2} \over m_{i} } + { F_{3} \over m_{f} }  ), 
\end{equation}
\begin{equation}f_2 =  {1 \over 2 } ( { F_{2} \over m_{i} } + { F_{3}
\over m_{f} }  ), 
\end{equation} 
\begin{equation}f_3 =  {1 \over 2 } ( {
F_{2} \over m_{i} } - { F_{3} \over m_{f} }  ); 
\end{equation}
\begin{equation}g_1 = G_1 - {1 \over 2 } ( \Delta m) ( { G_{2} \over
m_{i} } + { G_{3} \over m_{f} }  ), 
\end{equation} 
\begin{equation}g_2 =
{1 \over 2 } ( { G_{2} \over m_{i} } + { G_{3} \over m_{f} }  ),
\end{equation} 
\begin{equation}g_3 =  {1 \over 2 } ( { G_{2} \over m_{i}
} - { G_{3} \over m_{f} }  ), \label {(f6)}  
\end{equation} where
$\Delta m = m_i - m_f $.  Employing the nonrelativistic quark model
framework, Cheng and Tseng [14] have calculated these form factors at
maximum $q^2$, $${ f_{1}(q_{m}^{2}) \over N_{fi} } = 1 - { {\Delta m}
\over { 2 m_i }  } +  { { \Delta m } \over { 4 m_i m _f } } (1- { {\bar
\Lambda } \over { 2 m_f } }) (m_i + m_f - \eta  \Delta m) $$
\begin{equation} - { { \Delta m } \over { 8 m_i m _f } } { {\bar \Lambda
} \over {  m_Q } } (m_i + m_f + \eta  \Delta m), \label {(f7)}
\end{equation} 
\begin{equation}{ g_{1}(q_{m}^{2}) \over N'_{fi} } = 1 +
{ {\Delta m {\bar \Lambda}} \over { 4 } } (  { { 1 } \over { m_i m_q } } -  {
{1} \over { m_f m_Q }} ), \label {(f8)}  
\end{equation} 
where
$ \eta~=~N'_{fi}/N_{fi}$, ${\bar \Lambda} = m_f - m_q $ and 
$ q_{m}^{2} = (\Delta m)^2$ denotes the maximum $q^2$ transfer. 
$N_{fi}$ and $N'_{fi}$ are the flavor factors, 
\begin{equation}N_{fi} =
_{flavor-spin} <B_f| b_q^{\dagger} b_Q | B_i >_{flavour-spin},  \label
{(nfi)} 
\end{equation} 
\begin{equation}N'_{fi} =  _{flavour-spin} <B_f|
b_q^{\dagger} \sigma_{z}^Q b_Q | B_i >_{flavour-spin},  \label {(nfid)}
\end{equation} for the heavy quark $Q$ in the parent baryon $B_i$
transiting into the light quark $q$ in the daughter baryon $B_f$. 
$m_Q$ and $m_q$ denote masses of these heavy and light quarks respectively.
The light diquark present in the parent baryon behaves as spectator.
In the absence of a direct evaluation,  $q^2$ dependence of 
the baryonic form factors 
can be realized by assuming a pole dominance of the form, 
\begin{equation}f(q^{2}) = { {f(0)} \over { (1 - { { q^{2}} \over
{m_{V}^{2}} } )^{n}} }, \label {(d1)}  
\end{equation}
\begin{equation}g(q^{2}) = { {g(0)} \over { (1 - { { q^{2}} \over
{m_{A}^{2}} } )^{n}} },\label {(d2)}  
\end{equation} 
where $m_V$ and $m_A$ denote, respectively, 
pole masses of the vector meson and axial-vector meson having 
the quantum numbers of the current involved. Generally for the baryons,
one takes $ n = 2$. 

\par Upto the first order of parameterization, the
factorization amplitudes are given by
\begin{equation}A_{fac} = - { { G_F} \over {\sqrt{2}} } F_C f_P a_k (m_i
- m_f) f_{1}^{B_i B_f } (m_{P}^{2}),\label {(fpv)}  
\end{equation}
\begin{equation}B_{fac} =  { { G_F} \over {\sqrt{2}} } F_C f_P a_k (m_i
+ m_f) g_{1}^{B_i B_f } (m_{P}^{2}). \label {(fpc)}  
\end{equation}
$F_{C}$ is the CKM factor.  $a's$ are the two undetermined coefficients
 assigned to the effective charged current, $a_1,$ and the effective
 neutral current, $a_2$, parts of the weak Hamiltonian given in eq.(1). Values 
of these parameters can be related to the QCD coefficients as
\begin{equation} a_{1,2}=c_{1,2}+\zeta c_{2,1}, 
\end{equation} 
where  $\zeta = 1/N_{color}$. The values 
\begin{equation}
 a_1~~=~~1.26,~~~a_2~~=~~-0.51, 
\end{equation} 
give the best fit
 to the experimental data on charm meson decays corresponding to
 $\zeta \rightarrow 0$ [16].  In this approach, the quark currents of weak
 Hamiltonian are  considered as interpolating meson fields generating a
 $q \bar q$ state. The factorization contributions, being proportional 
to the meson momenta, can be considered as the correction to the decay amplitudes
obtained in the CA framework which employs the soft meson limit.
\subsubsection {ETC and Pole Contributions} 
The second term $M_{ETC}$ in
eq(8), corresponding to  the equal time current commutator (ETC), is
given by the matrix element of $H_W$ between the initial  and the final
state baryons,
\begin{equation} \langle B_{f} \vert H_{W}\vert B_{i}
\rangle = \bar u_{f}(P_f) ( a_{if} - b_{if} \gamma_{5} ) u_{i} (P_i).
\label {(me)} 
\end{equation} 
It is well known that the PV matrix
elements $b_{if}$ vanish for the hyperons due to C-parity null
theorem [9] in the flavor symmetry limit. In the case of the charm baryon 
decays, in analogy with hyperons, it has been shown that $b_{if} << a_{if} $.
Hence, the ETC term enters only in the s-wave (PV) amplitudes;
\begin{equation}
A^{ETC} = { {1} \over {f_{k} }}  < B_f | [ Q_{k}^{5} ,
H_{w}^{PV} ] | B_i > = { {1} \over {f_{k} }}  < B_f | [ Q_{k},
H_{w}^{PC} ] | B_i >, \label {(ETC)} 
\end{equation} 
where $Q_k$ and
$Q_{k}^5 $ denote the vector and axial vector charges respectively.  The
p - wave (PC) decay amplitudes are then described by the $J^P = 1/2^{+}
$ pole terms $(M_{pole})$. The baryon  pole terms, arising from s- and u-
channels contributions to PC decay amplitude, are given by
\begin{equation}
B_{pole} = {  { g_{ \ell fk} a_{i \ell} } \over { m_i -
m_{\ell} } } {  { m_i + m_f } \over { m_{\ell }+ m_{f} } } + {  { g_{i
\ell 'k} a_{\ell' f} } \over { m_f - m_{\ell '} } } {  { m_i + m_f } \over {
m_{i}+ m_{\ell '} } }, \label {(pole)} 
\end{equation} 
where $g_{ijk}$ are
the strong baryon-baryon coupling constants, $\ell$ and $\ell '$ are the
intermediate states - corresponding to the respective s- and u-channels. This
pole contribution differs from the simple pole model calculations due to
the appearance of extra mass factors. This term is actually a modified
pole term and contains the contributions from the surface term, the soft-meson
Born-term contraction and the baryon-pole term [9], combined in a well-defined 
way. It has been pointed out by Karlsen
and Scadron [10] that in this way this term accounts for the large
momentum dependence away from the soft pion limit as occurs in the
charmed baryon decays. The weak matrix elements $a_{ij}$ for baryonic
transition $B_i \rightarrow B_f $ are evaluated in the constituent quark
model following the work of Riazuddin and Fayyazuddin [17]. For the
strong baryon-meson coupling constants $ g_{ijk}$, we introduce SU(4)
breaking effects [18] through
\begin{equation}
g_{B B' P} = { { M_B + M_{B'}
}\over { 2 M_N } } g_{B B' P}^{sym} \label {(gbb)},
\end{equation} 
where $ g_{B B' P}^{sym} $ denotes the SU(4) symmetric coupling.
\section {Numerical Calculations and Discussion of Results}
We first determine the factorizable contributions to the Cabibbo-angle enhanced
decays of $\Lambda_c^+$ baryon using the HQET guided form factors, which
have been calculated earlier in the nonrelativistic quark model framework [14];
\begin{equation}f_{1}^{ \Lambda_c \Lambda } = 0.50 N_{ \Lambda_c \Lambda
}, ~~~g_{1}^{ \Lambda_c \Lambda } = 0.65 N_{ \Lambda_c \Lambda };
\end{equation} 
\begin{equation}
f_{1}^{ \Lambda_c p } = 0.34 N_{
\Lambda_c p }, ~~~g_{1}^{ \Lambda_c p } = 0.53 N_{ \Lambda_c p },
\end{equation} 
where the flavor-spin factors are
\begin{equation} 
N_{ \Lambda_c \Lambda}' ~=~ N_{ \Lambda_c \Lambda}~~ = 
~~{{1} \over {\sqrt 3}};~~~
N_{ \Lambda_c p }' ~~=~~N_{ \Lambda_c p}~~ = ~~{{1} \over {\sqrt 2}}. 
\end{equation}
Reliability of these form factors has been well tested by
computing decay width of the semileptonic mode $ \Lambda_c^+ \rightarrow
\Lambda e^+ \nu_e $, 
\begin{equation}
\Gamma (\Lambda_c^+ \rightarrow
\Lambda e^+ \nu_e) = 7.1 \times 10^{10} s^{-1} 
\end{equation} 
which is consistent with the experimental value [1]. It is worth pointing
that the flavor factors $N_{\Lambda_c \Lambda } $
plays a crucial role for the agreement. Earlier theoretical models [19]  
have given quite large values for this semileptonic decay rate. The weak
Hamiltonian eq(1) allows only $ \Lambda_c^+
\rightarrow p\bar {K^0}/\Lambda\pi^+$ decays to receive the factorization
contributions. For these decays, the form factors given in eqs. (35) and (36)
yield the following branching ratios and asymmetries:
\begin{equation} Br(\Lambda_c^+\rightarrow p\bar {K^0}) = 0.48\% ~~
((2.2\pm0.4)\% ~Expt.), 
\end{equation}
\begin{equation} \alpha(\Lambda_c^+\rightarrow p\bar {K^0}) = - 0.94,
\end{equation}
 \begin{equation} Br(\Lambda_c^+\rightarrow \Lambda \pi^+) = 1.29\%~~
((0.79 \pm 0.18)\% ~~Expt.), 
\end{equation}
 \begin{equation} \alpha(\Lambda_c^+\rightarrow \Lambda \pi^+) = - 0.97~~
(-0.94 \pm 0.29 ~~Expt.). 
\end{equation}
Though, the asymmetry of $\Lambda_c^+\rightarrow \Lambda \pi^+$ is in good
agreement with experiment, its branching ratio is rather large. In contrast, 
branching ratio of $\Lambda_c^+\rightarrow p\bar {K^0}$, is much less than the
experimental value. Thus, the spectator contributions alone cannot
explain even these decays.
The branching ratio of $\pi^+$ emitting mode is greater than that of the 
$\bar {K^0}$ emitting mode due to the color enhancement factor $(a_1/a_2)^2$. 
However, in a typical $\pi^+$ emitting decay
$\Lambda_c^+\rightarrow \Sigma^0 \pi^+,$ the factorization term vanishes
due to the Clebsch-Gordon coefficient. It proceeds only through the
nonspectator processes, which are also responsible for the 
remaining $\Lambda_c^+$ decays where the factorization terms do not appear.
Accurate  experimental measurements of these
decays can clearly determine the relative strength of the nonspectator terms in
the charmed baryon decays. 

\par In our framework, nonspectator ETC and pole terms
involve the matrix elements of the kind
 $\langle B_{f} \vert H_{W}^{PC} \vert B_{i} \rangle$. We evaluate  such
matrix elements following the scheme of Riazuddin and Fayyazuddin [17], which
gives the nonrelativistic reduction of the PC-Hamiltonian, 
\begin{equation} H_W^{PC} =
c_-(m_c)(s^{\dag}c~u^{\dag}d - s^{\dag}{\bf \sigma} c \cdot u^{\dag} {\bf
\sigma} d)\delta^3({\bf r}). 
\end{equation} 
Note that only $c_-$ appears
in this limit, because the part of Hamiltonian corresponding to $c_+$
 is symmetric in the color indices. We take the QCD enhancement at
the charm mass scale $c_-(m_c) = 1.75,$ which is lower than $c_-(m_s) = 2.23$
used in the hyperon sector. To reduce the number of free parameters, we 
determine the scale for the ETC and pole terms using
\begin{equation} \langle \psi_{\Lambda}
\vert \delta^3(r)\vert \psi_{\Lambda_c^+} \rangle \approx \langle
\psi_{p} \vert \delta^3(r)\vert \psi_{\Sigma^+} \rangle. 
\end{equation}

\par Combining all the ingredients of PV and PC decay
amplitudes, we compute the branching ratios and asymmetries for various
decays.  These are given in the Table 1.
Experimentally measured [1] masses, lifetimes,  and decay constants have
been used in the present analysis. Comparing the theoretical values with 
those obtained in the
pure factorization case, we  find that inclusion of the nonspectator
terms modifies the branching ratios in the desired direction without affecting
the asymmetry parameters. We make the following observations: 

\par \noindent {\bf 1.} The branching fraction for $\Lambda_c^+\rightarrow
p\bar{K^0}$ increases from $0.48\%$ to $1.23\%$ bringing it closer to
the experiment.  The increase in the branching ratio
occurs due to constructive interference between the ETC and
factorization terms, comparable in magnitude, in the PV mode. Similarly
for the PC mode also, the pole and factorization terms interfere
constructively, though the pole contribution is around $30\%$ only. 
We predict its asymmetry $\alpha(\Lambda_c^+\rightarrow
p\bar{K^0})$~ =~- 0.99.
\par
\noindent {\bf 2.} For the decay $\Lambda_c^+\rightarrow \Lambda \pi^+$,
the branching ratio  decreases from $1.29\%$ to $1.17\%$ in the right
direction. For this decay, the ETC contribution vanishes, so its PV
amplitude is given only by the factorization term. For its PC amplitude,
there exists a destructive interference between the pole and
factorization contributions for the choice of the form factors given in
eq. (35). We wish to remark that even if the pole and factorization
terms interfere constructively, its branching ratio 
would hardly be raised to 1.44\%. This is due to the reason that the pole
terms in s- and u-channels tend to cancel each  other thereby reducing
the pole strength to around $10\%$ of the factorization. We obtain its
asymmetry $\alpha (\Lambda_c^+ \rightarrow \Lambda \pi^+) ~=~~- 0.99$ in
nice agreement with the experimental value recently measured by the CLEO
collaboration. The CLEO measurement [3] has determined the following
sets of PV and PC amplitudes (in the units of $~G_F V_{ud} V_{cs}^*
\times 10^{-2}~ GeV^2)$:
\begin{equation} Set I:~~~  A(\Lambda_c^+\rightarrow    \Lambda \pi^+)~=~
-3.0_{-1.2}^{+0.8}, ~~~
B(\Lambda_c^+\rightarrow   \Lambda \pi^+)~=~+12.7_{- 2.5}^{+2.7}~; 
\end{equation} 
\begin{equation} Set II: ~~~
A(\Lambda_c^+\rightarrow    \Lambda \pi^+)~=~
 -4.3_{-0.9}^{+0.8}, ~~~
B(\Lambda_c^+\rightarrow   \Lambda \pi^+)~=~
+8.9_{-2.4}^{+3.4}. 
\end{equation} 
Our analysis gives 
\begin{equation}
A(\Lambda_c^+\rightarrow \Lambda \pi^+)~=~-4.6,~~
B(\Lambda_c^+\rightarrow \Lambda \pi^+)~=~+15.8,~~
\end{equation} 
which seem to favor the first set. As this decay occurs largely through the
spectator quark process, the present data seems to demand lower values
of the form factors involved, or more accurate measurement is desired to
clarify the situation. 

\par \noindent {\bf 3.} The same CLEO experiment
[3] has measured the asymmetry of $\Lambda_c^+\rightarrow \Sigma^+
\pi^0$ decay, 
\begin{equation} \alpha(\Lambda_c^+\rightarrow \Sigma^+
\pi^0)~=~ -0.45\pm 0.31,
\end{equation} which is in good agreement with
our prediction, 
\begin{equation} \alpha(\Lambda_c^+\rightarrow \Sigma^+
\pi^0)~=~ -0.31.
\end{equation} In contrast, earlier theoretical efforts
[6-8] have given large positive value, ranging from 0.78 to 0.92, for
this asymmetry parameter. The calculated branching ratio in our
analysis, 
\begin{equation} Br(\Lambda_c^+\rightarrow \Sigma^+ \pi^0) =
0.69\% ~~((0.88 \pm 0.22)\% ~~Expt.), 
\end{equation} also matches well
the experimental value. Considering the PV and PC amplitudes explicitly,
the measured values are (in the units of $~G_F V_{ud} V_{cs}^*
\times 10^{-2}~ GeV^2)$;
\begin{equation} Set I:~~~  A(\Lambda_c^+\rightarrow
\Sigma^+    \pi^0)~=~+1.3_{- 1.1}^{+0.9} , ~~~
B(\Lambda_c^+\rightarrow   \Sigma^+   \pi^0)~=~-17.3_{- 2.9}^{+2.3};
\end{equation} 
\begin{equation}  Set II:~~~  A(\Lambda_c^+\rightarrow
\Sigma^+    \pi^0)~=~
+5.4_{-0.7}^{+0.9}, ~~~
B(\Lambda_c^+\rightarrow   \Sigma^+   \pi^0)~=~~ -4.1_{-3.0}^{+3.4}. 
\end{equation} 
For these decay amplitudes, we
obtain, 
\begin{equation} A(\Lambda_c^+\rightarrow \Sigma^+
\pi^0)~=~+5.4;~~ B(\Lambda_c^+\rightarrow \Sigma^+
\pi^0)~=~-2.7,~~
\end{equation} 
consistent with the second set. 

\par \noindent {\bf 4.} For $\Lambda_c^+\rightarrow \Sigma^0 \pi^+ $ decay,
our analysis yields, 
\begin{equation} Br(\Lambda_c^+\rightarrow \Sigma^0
\pi^+) = 0.69\%~~ ((0.88 \pm 0.20)\% ~~Expt.),
\end{equation} agreeing
well with the experiment, and the asymmetry
$\alpha(\Lambda_c^+\rightarrow \Sigma^0 \pi^+) = -0.31$, i.e. the same
as that of the $\Lambda_c^+\rightarrow \Sigma^+ \pi^0$, as expected from
the isospin symmetry arguments. 
\par \noindent {\bf 5.} For $\eta-\eta'$
emitting decays, we calculate: 
\begin{equation}
Br(\Lambda_c^+\rightarrow   \Sigma^+   \eta) ~=~0.26\% ~~ ((0.48\pm
0.17)\% ~~Expt.),
\end{equation}
\begin{equation}\alpha(\Lambda_c^+\rightarrow    \Sigma^+    \eta) ~=~-
0.99;
\end{equation} 
\begin{equation} Br(\Lambda_c^+\rightarrow \Sigma^+
\eta') ~=~0.08\%,
\end{equation}
\begin{equation}\alpha(\Lambda_c^+\rightarrow \Sigma^+        \eta')
~=~+0.49; 
\end{equation} 
for the $\eta-\eta'$ physical mixing angle
$-10^o$. Here, the branching ratio of 
$\Lambda_c^+\rightarrow   \Sigma^+   \eta $ decay is consistent with the
observed value. We find that this branching ratio comes closer
($0.29\%$) to the experiment for the physical mixing angle $(-23^o)$   
given by the linear mass formulae [1].

\par \noindent {\bf 6.} The decay $\Lambda_c^+\rightarrow \Xi^0K^+,$ is
theoretically the cleanest of all the $\Lambda_c^+$ decays as it acquires only p-wave
contribution to its decay amplitude and has null asymmetry. For this
mode, we obtain 
\begin{equation} Br(\Lambda_c^+\rightarrow \Xi^0
K^+)~~=~~0.07\%~~((0.34\pm 0.09)\%~~ Expt.),
\end{equation} 
which is smaller than the experimental value. 
\par \noindent {\bf 7.} Among the $\Xi_c^+$ decays, there are only two
possible modes. Recently, the branching ratio of  $\Xi^+_c\rightarrow
\Xi^0 \pi^+$ decay has also been measured in a CLEO experiment [2], for which 
our analysis yields,
\begin{equation} Br(\Xi_c^+\rightarrow \Xi^0
\pi^+)~~=~~1.08\% ~~((1.2\pm 0.5\pm 0.3)\% ~~Expt.),
 \end{equation}
in excellent agreement with experiment.
It may be remarked that though, both the  $\Xi_c^+$ modes get
contributions from the factorization, pole, and ETC terms, yet the decay
$(\Xi^+_c\rightarrow \Xi^0 \pi^+)$ dominates over $(\Xi^+_c\rightarrow
\Sigma^+ \bar {K^0}) $ by an order of magnitude
\begin{equation} \frac {Br(\Xi^+_c\rightarrow \Xi^0 \pi^+)}
{Br(\Xi^+_c\rightarrow \Sigma^+ \bar {K^0})} ~=~13.2. 
\end{equation}

\par \noindent {\bf 8.} Among $\Xi_c^0$ decays, we find that
the dominant mode is $\Xi^0_c\rightarrow \Xi^- \pi^+~$ which has branching 
ratio around 2\% in our model.

\subsection {Variation of $|\psi(0)|^2$}
So far, we have taken the scale $|\psi(0)|^2$ for the nonspectator
terms same as that of the hyperon sector.
However, this being a dimensional  quantity, it may be incorrect to ignore
its variation with flavor. Unfortunately, evaluation of $|\psi(0)|^2$ is
as yet uncertain for the baryons and more complicated, because unlike mesons
these are three body systems. However, a naive estimate for the scale
may be obtained using the hyperfine splitting,
\begin{equation} \Delta E_{HFS} = { {4\pi\alpha_s } \over { 9m_1m_2 } }
\vert   \psi(0)\vert^2   \langle{\bf   \sigma_1} \cdot {\bf   \sigma_2}   \rangle, 
\end{equation}
leading to 
\begin{equation} {{\Sigma_c - \Lambda_c} \over {\Sigma -
\Lambda}} ={ {\vert \psi(0)\vert^2_c} \over {\vert \psi(0)\vert^2_s}}
 {{\alpha_s(m_c)} \over {\alpha_s(m_s)}}  {{m_s(m_c-m_u)} \over
{m_c(m_s-m_u)}}. 
\end{equation} 
For the choice
$\alpha_s(m_c)/\alpha_s(m_s) \approx 0.53,$ we obtain, $r ~\equiv ~ \vert
\psi(0)\vert^2_c /\vert \psi(0)\vert^2_s \approx 2.1$. However, we do not
expect this ratio to hold for the weak decays considered in the present work,
as the weak baryon transitions occurring in the charmed baryon decays 
involve  $~ _s< \psi~ | \delta^3({\bf r}) |~ \psi >_c ~$ which should lie
between 1 and 2. We have investigated the implications of this 
scale ratio, varying from 1 to 2,  on 
the branching ratios and asymmetry parameters. 
We make the following observations:

\par  \noindent  {\bf  1.}  Asymmetry of  all  the
decays, except those of $\Xi_c \rightarrow \Sigma + {\bar K}^0$, 
remain almost unaffected and stay in good agreement with the experiment.  
Asymmetry of the $\Xi_c \rightarrow \Sigma + {\bar K}^0$ decays show
change in sign for scale parameter is increased.

\par \noindent {\bf 2.} Branching ratios of $\Lambda_c^+\rightarrow p \bar
{K^0}/\Lambda \pi^+/\Xi^0 K^+$ decays are found to require this ratio
on the higher side (1.5 to 2.1) for better agreement with the
experiment, whereas the $\Lambda_c^+ \rightarrow \Sigma \pi / \eta $ decays
prefer a small enhancement ratio (1.1 to 1.3).

\par \noindent {\bf 3.} Ratio of the decay rates,
\begin{equation}
{{ Br(\Lambda_c^+\rightarrow \Lambda\pi^+) } \over
{ Br(\Lambda_c^+\rightarrow p\bar {K^0}) }  } ~= ~(0.92 ~to ~0.40), 
\end{equation} 
for the chosen range $(~ r~ = ~1~ to~ 2~)$ approaching 
the experimental value ${0.36\pm 0.10}$. It may be noticed that this ratio
has been theoretically estimated to be as high as 13 in some of the 
earlier models due to the expected color enhancement.

\par \noindent {\bf 4.} The scale ratio certainly increases the
 branching ratio of $\Lambda_c^+\rightarrow \Xi^0 ~K^+$ as desired by the
experiment.  
Since all the decays $\Lambda_c^+\rightarrow
\Sigma^+\pi^0/\Sigma^0\pi^+$, 
$\Lambda_c^+ \rightarrow \Sigma^+\eta/\Sigma^+\eta'$
 and $\Lambda_c^+ \rightarrow \Xi^0K^+$ occur
only through the nonspectator terms, their
relative ratios remain independent of the scale factor in our analysis. 

\par \noindent {\bf 5.}
We expect the ratio $r$ to lie close to 1.4 using the following ansatz:
\begin{equation}
(~ _s< \psi | \delta^3({\bf r}) |  \psi >_c~)^2 ~~\approx~~
_s< \psi | \delta^3({\bf r}) |  \psi >_s ~  \times ~
_c< \psi | \delta^3({\bf r}) | \psi >_c .
\end{equation}
To show the trends of the results, we give corresponding values of the 
branching ratios and asymmetry parameters of the charmed baryon decays 
in the Table 2.  

\section{Summary and Conclusions}
We have studied the two-body Cabibbo-angle favored decays of the charmed
baryons $\Lambda_c^+$, $\Xi_c^+$, and $\Xi_c^0$ into the octet baryons
 and pseudoscalar mesons. It is now established that factorization alone
 cannot explain the available data as the branching ratio of
 $\Lambda_c^+\rightarrow \Sigma \pi/\eta$ and 
 $\Lambda_c^+\rightarrow \Xi^0 K^+$ decays,
 forbidden in the factorization scheme,
 have been measured to be comparable to that of $\Lambda _c^+ \rightarrow
 \Lambda \pi^+$. Hence the nonspectator processes, like W-exchange diagram,
 seem to play
 a significant role in understanding these decays. Further,
 $\Lambda_c^+\rightarrow p {\bar K}^0$  decay which receives color-
suppressed factorization has branching ratio greater than that of the
 color-favored decay  $\Lambda _c^+ \rightarrow \Lambda \pi^+$
by a factor of 2.5. 

\par In the absence of a direct method for calculating the
nonspectator terms, we have employed the standard
current algebra framework to estimate their strength. The factorization 
contributions, being proportional to the meson momentum, provide
 corrections to this framework. We have evaluated the
factorization terms using the HQET guided baryonic form factors.
We have obtained branching ratios and asymmetry parameters for these
decays, which are found to be consistent with the experimental data.
We observe that inclusion of the 
nonspectator contributions increases $Br(\Lambda_c^+\rightarrow p\bar{K^0})$
from $0.48\%$ to $1.23\%$ and decreases  $Br(\Lambda_c^+\rightarrow
\Lambda \pi^+)$ from $1.29\%$ to $1.17\%$  in the desired directions.
Further, branching ratios of $\Lambda_c^+\rightarrow \Sigma^+
\pi^0/\Sigma^0\pi^+/ \Sigma^+\eta $ and the only measured 
$\Xi^+_c\rightarrow \Xi^0\pi^+$ decay, obtained in the present analysis, are in
good agreement with the experiment. 
The experimentally available asymmetries of 
$\Lambda^+_c$ decays are also found in nice agreement with our results.
However, branching ratio of $\Lambda_c^+ \rightarrow \Xi^0 K^+$ decay is
found to be much less than the observed value. Theoretically, one expect
$\Lambda^+_c\rightarrow \Xi^0 K^+$ to be the cleanest of all the modes as
neither factorization nor ETC term contributes to this process, so it should 
provide a measure of the pole terms. 
\par We have also investigated 
the effects, flavor dependence
of $|\psi(0)|^2$ as is evident by $\Sigma_c$ and $\Lambda_c$ mass splitting,
on these decays. It
can result into the desired enhancement of $\Lambda^+_c\rightarrow
\Xi^0K^+$ but simultaneously it would increase the decay rates of 
$\Lambda_c^+\rightarrow
\Sigma \pi/ \eta $, as these bear fixed ratios in our model.
Though we find that a small scale enhancement is acceptable to  
the present level of data on charmed baryon decays, it needs some new
physics. Small branching ratio for $\Lambda \rightarrow \Xi^0 K^+$ decay
 in fact results due to near cancellation of the pole terms in the s- 
and u- channels, which involve antitriplet (C = 1) to octet baryon and
sextet (C = 1) to octet baryon transition respectively. We expect that
the HQET considerations may differentiate between the two type of the
transitions. Further, 
 final state interactions (FSI), well known to substantially alter the 
decay rates of the charm mesons, may also affect the charm baryon decays
by feeding one decay mode into the other.

\vskip 1.0 cm 
\noindent {\bf Acknowledgements}\\ 
One of the authors (RCV) gratefully acknowledges the financial support from
the Department of Science and Technology, New Delhi, India.

\newpage
\begin{table} \begin{center} \caption {Branching Ratios and Asymmetries
of Charmed Baryon Decays ($~r = 1$)} \vskip 0.3 cm
\begin{tabular}{|c|c|c|c|c|} \hline Decay  &  Br.(\%) & Expt. Br.
(\%)  & $\alpha$ & Expt. $\alpha$ \\ \hline \hline $
\Lambda_{c}^{+} \rightarrow p \bar K^{0}$ &  $1.23$ & $2.2\pm 0.4$  &
$-0.99$ &  \\ $ \Lambda_{c}^{+}\rightarrow \Lambda \pi^{+}$ &  $ 1.17$ &
$0.79 \pm 0.18$ &  $-0.99$ & $ -0.94 \pm 0.24$ \\ $
\Lambda_{c}^{+}\rightarrow \Sigma^{+} \pi^{0} $ & $0.69$ & $0.88 \pm
0.22$ &   $-0.31 $ & $-0.45 \pm 0.31$ \\ $ \Lambda_{c}^{+}\rightarrow
\Sigma^{+} \eta $ & $ 0.26^a~(0.29^b)$ & $0.48 \pm 0.17$ &  $-0.99^a~(-0.91^b)$ & \\ $
\Lambda_{c}^{+}\rightarrow \Sigma^{+} \eta' $ & $ 0.08^a~(0.05^b)$ &  & $+0.49^a~(+0.78^b)$ &
\\  $\Lambda_{c}^{+}\rightarrow \Sigma^{0} \pi^{+} $     &$ 0.69$
&$0.88\pm 0.20$ & $-0.31$ & \\ $ \Lambda_{c}^{+} \rightarrow \Xi^{0}
K^{+}$ & $ 0.07$   &$0.34 \pm 0.09$ & $ 0.00$ & \\  &   &   &   &  \\ $
\Xi_{c}^{+} \rightarrow \Xi^{0} \pi^{+}$ & $ 1.08 $ &  $ 1.2\pm0.5\pm0.3
$ &  $-0.74$ &   \\ $ \Xi_{c}^{+} \rightarrow \Sigma^{+} \bar K^{0}$  &
$ 0.08$ &  &  $-0.38$ &   \\ &   &   &   &  \\ $ \Xi_{c}^{0} \rightarrow
\Xi^{0} \pi^{0}$  & $ 0.44$ &  &  $-0.80$ &    \\ $ \Xi_{c}^{0}
\rightarrow \Xi^{0} \eta$ & $0.08^a~(0.11^b)$ &   &  $+0.01^a~(+0.21^b)$ &     \\ $ \Xi_{c}^{0}
\rightarrow \Xi^{0} \eta'$ & $0.05^a~(0.03^b)$ &   &  $+0.68^a~(+0.80^b)$ &     \\ $
\Xi_{c}^{0} \rightarrow \Xi^{-} \pi^{+}$ & $ 1.99$ &   &  $-0.99$ &
\\ $ \Xi_{c}^{0} \rightarrow \Sigma^{+} K^{-}$ & $ 0.06$ &  & $0.00 $ &
\\ $ \Xi_{c}^{0} \rightarrow \Sigma^{0} \bar K^{0}$ & $ 0.08$ &   &
$-0.15 $ &    \\ $ \Xi_{c}^{0} \rightarrow \Lambda \bar K^{0}$ & $0.34$
&   &  $-0.85$ & \\ \hline
\end{tabular}  $a~ =~\phi_{\eta-\eta'}~ = ~-10^o,~~~b~
=~\phi_{\eta-\eta'}~ = ~-23^o$
 \end{center} \end{table}

\newpage
\begin{table} \begin{center} \caption{Branching Ratios and Asymmetries
of Charmed Baryons ($~r = 1.4$)} \vskip 0.3 cm
\begin{tabular}{|c|c|c|c|c|} \hline Decay  &  Br. (\%) & Expt. Br.
(\%)  & $\alpha$  & Expt. $\alpha$ \\ \hline \hline $
\Lambda_{c}^{+} \rightarrow p \bar K^{0}$ &  $1.64$ & $2.2\pm 0.4$  &
$-0.98$ &  \\ $ \Lambda_{c}^{+}\rightarrow \Lambda \pi^{+}$ &  $ 1.12$ &
$0.79 \pm 0.18$ &  $-0.99$ & $ -0.94 \pm 0.24$ \\ $
\Lambda_{c}^{+}\rightarrow \Sigma^{+} \pi^{0} $ & $1.34$ & $0.88 \pm
0.22$ &   $-0.31 $ & $-0.45 \pm 0.31$ \\ $ \Lambda_{c}^{+}\rightarrow
\Sigma^{+} \eta $ & $ 0.50^a~(0.57^b)$ & $0.48 \pm 0.17$ &  $-0.99^a~(-0.91^b)$ & \\ $
\Lambda_{c}^{+}\rightarrow \Sigma^{+} \eta' $ & $ 0.15^a~(0.10^b)$ &  & $+0.49^a~(+0.78^b)$ &
\\  $\Lambda_{c}^{+}\rightarrow \Sigma^{0} \pi^{+} $     &$ 1.34$
&$0.88\pm 0.20$ & $-0.31$ & \\ $ \Lambda_{c}^{+} \rightarrow \Xi^{0}
K^{+}$ & $ 0.13$   &$0.34 \pm 0.09$ & $ 0.00$ & \\  &   &   &   &  \\ $
\Xi_{c}^{+} \rightarrow \Xi^{0} \pi^{+}$ & $ 0.53 $ &  $ 1.2\pm0.5\pm0.3
$ &  $-0.27$ &   \\ $ \Xi_{c}^{+} \rightarrow \Sigma^{+} \bar K^{0}$  &
$ 0.04$ &  &  $+0.54$ &   \\ &   &   &   &  \\ $ \Xi_{c}^{0} \rightarrow
\Xi^{0} \pi^{0}$  & $ 0.87$ &  &  $-0.80$ &    \\ $ \Xi_{c}^{0}
\rightarrow \Xi^{0} \eta$ & $0.16^a~(0.22^b)$ &   &  $+0.01^a~(+0.21^b)$ &     \\ $ \Xi_{c}^{0}
\rightarrow \Xi^{0} \eta'$ & $0.10^a~(0.06^b)$ &   &  $+0.68^a~(+0.80^b)$ &     \\ $
\Xi_{c}^{0} \rightarrow \Xi^{-} \pi^{+}$ & $ 2.46$ &   &  $-0.97$ &
\\ $ \Xi_{c}^{0} \rightarrow \Sigma^{+} K^{-}$ & $ 0.12$ &  & $0.00 $ &
\\ $ \Xi_{c}^{0} \rightarrow \Sigma^{0} \bar K^{0}$ & $ 0.07$ &   &
$+0.48 $ &    \\ $ \Xi_{c}^{0} \rightarrow \Lambda \bar K^{0}$ & $0.54$
&   &  $-0.79$ & \\ \hline  
\end{tabular} $a~ =~\phi_{\eta-\eta'}~ = ~-10^o,~~~b~
=~\phi_{\eta-\eta'}~ = ~-23^o$
\end{center} \end{table}

\end{document}